\documentclass[zpreprint,zbstpl]{./zeus_paper}
\usepackage[english]{babel}

\newcommand{\ZcoosysB}{%
The ZEUS coordinate system is a right-handed Cartesian system, with the $Z$
axis pointing in the proton beam direction, referred to as the ``forward
direction'', and the $X$ axis pointing left towards the centre of HERA.
The coordinate origin is at the nominal interaction point.\xspace}

\newcommand{\ZcoosysfnB}{\footnote{\ZcoosysB}}

\chardef\usc=95
\chardef\til=126
\catcode`\@=11 % @ signs are now treated as letters
\DeclareRobustCommand\xdotspace{\futurelet\@let@token\@xdotspace}
\def\@xdotspace{%
  \ifx\@let@token.\else
  \ifx\@let@token\bgroup.\else
  \ifx\@let@token\egroup.\else
  \ifx\@let@token\/.\else
  \ifx\@let@token\ .\else
  \ifx\@let@token~.\else
  \ifx\@let@token!.\else
  \ifx\@let@token,.\else
  \ifx\@let@token:.\else
  \ifx\@let@token;.\else
  \ifx\@let@token?.\else
  \ifx\@let@token/.\else
  \ifx\@let@token'.\else
  \ifx\@let@token).\else
  \ifx\@let@token-.\else
  \ifx\@let@token\@xobeysp.\else
  \ifx\@let@token\space.\else
  \ifx\@let@token\@sptoken.\else
   .\space
   \fi\fi\fi\fi\fi\fi\fi\fi\fi\fi\fi\fi\fi\fi\fi\fi\fi\fi}
\catcode`\@=12 % @ signs are no longer letters

\newcommand{\stru}[2]{%
   \relax\ifmmode\hbox{\vrule height#1 depth#2 width0pt}%
   \else\vrule height#1 depth#2 width0pt\fi}
\newcommand{\Ronum}[1]{\uppercase\expandafter{\romannumeral#1}}
\newcommand{\ronum}[1]{\expandafter{\romannumeral#1}}
\DeclareRobustCommand{\LaTeXZ}{%
  \LaTeX\kern-.05em4\kern-.1em
  {\raisebox{-0.2ex}{$\scriptstyle\text{ZEUS}$}}\xspace}
\DeclareMathAlphabet{\mathbf}{OT1}{cmr}{bx}{sl}
\newcommand{\eVdist}{\kern-0.06667em}

\newcommand{\Tesla}{\,\text{T}}

\newcommand{\slashfrac}[2]{%
  \raisebox{0.5ex}{\ensuremath #1}\kern-0.12em/\kern-0.08em
  \raisebox{-.8ex}{\ensuremath #2}}
\newcommand{\sqr}[3]{%
    {\vcenter{\hrule height.#3ex\hbox{\vrule width.#2ex height#1ex
     \kern#1ex\vrule width.#3ex}\hrule height.#2ex}}}

\catcode`\@=11 % @ signs are now treated as letters
\newcommand{\parenbar}{\mathpalette\p@renb@r}
\def\p@renb@r#1#2{\vbox{%
  \ifx#1\scriptscriptstyle \dimen@.7em\dimen@ii.2em\else
  \ifx#1\scriptstyle \dimen@.8em\dimen@ii.25em\else
  \dimen@1em\dimen@ii.4em\fi\fi \offinterlineskip
  \ialign{\hfill##\hfill\cr
    \vbox{\hrule width\dimen@ii}\cr
    \noalign{\vskip-.3ex}%
    \hbox to\dimen@{$\mathchar300\hfil\mathchar301$}\cr
    \noalign{\vskip-.3ex}%
    $#1#2$\cr}}}
\catcode`\@=12 % @ signs are no longer letters

\newcommand{\rnge}{\hbox{$\,\text{--}\,$}}

\newcommand{\IP}{{\rm I$\kern-0.01667em$P}\xspace}

\mathchardef\qsm=63
\mathchardef\pls=43
\mathchardef\mns=512
\mathchardef\plm=518
\mathchardef\eql=61
\mathchardef\smallleft=300
\mathchardef\smallright=301
\mathchardef\les=316
\mathchardef\gre=318
\mathchardef\leq=532
\mathchardef\grq=533
\catcode`\@=11 % @ signs are now treated as letters
\newcounter{pict@width}
\newcounter{pict@height}
\newlength{\pict@scale}
\newcommand{\psfigadd}[4]{%
\setcounter{pict@width}{1*\ratio{#2+\pict@scale/2}{\pict@scale}}
\setcounter{pict@height}{1*\ratio{#3+\pict@scale/2}{\pict@scale}}
\setlength{\unitlength}{\pict@scale}
\hbox to #2{\hspace{-\fill}\begin{picture}(\thepict@width,\thepict@height)
\put(0,0){\psfig{figure=#1,width=#2,height=#3,clip=}}
\SetScale{0.283466457}
\SetWidth{1.763889}
{#4}
\end{picture}}
}
\newcounter{pict@widthfst}
\newcounter{pict@widthscd}
\newcounter{pict@widthtot}
\newcommand{\psfigaddtwo}[7]{%
\setcounter{pict@widthfst}{1*\ratio{#2+\pict@scale/2}{\pict@scale}}
\setcounter{pict@widthscd}{1*\ratio{#2+#4+\pict@scale/2}{\pict@scale}}
\setcounter{pict@widthtot}{1*\ratio{#2+#4+#6+\pict@scale/2}{\pict@scale}}
\setcounter{pict@height}{1*\ratio{#3+\pict@scale/2}{\pict@scale}}
\setlength{\unitlength}{\pict@scale}
\hbox{\hspace{-\fill}\begin{picture}(\thepict@widthtot,\thepict@height)
\put(0,0){\psfig{figure=#1,width=#2,height=#3,clip=}}
\put(\thepict@widthscd,0){\psfig{figure=#5,width=#6,height=#3,clip=}}
\SetScale{0.283466457}
\SetWidth{1.763889}
{#7}
\end{picture}}
}
\newcommand{\psfigror}[4]{%
\setcounter{pict@width}{1*\ratio{#2+\pict@scale/2}{\pict@scale}}
\setcounter{pict@height}{1*\ratio{#3+\pict@scale/2}{\pict@scale}}
\setlength{\unitlength}{\pict@scale}
\hbox{\begin{picture}(\thepict@width,\thepict@height)
\put(0,\thepict@height){\psfig{figure=#1,width=#3,height=#2,clip=,angle=270}}
\SetScale{0.283466457}
\SetWidth{1.763889}
{#4}
\end{picture}}
}
\newcommand{\psfigrol}[4]{%
\setcounter{pict@width}{1*\ratio{#2+\pict@scale/2}{\pict@scale}}
\setcounter{pict@height}{1*\ratio{#3+\pict@scale/2}{\pict@scale}}
\setlength{\unitlength}{\pict@scale}
\hbox{\begin{picture}(\thepict@width,\thepict@height)
\put(0,0){\psfig{figure=#1,width=#3,height=#2,clip=,angle=90}}
\SetScale{0.283466457}
\SetWidth{1.763889}
{#4}
\end{picture}}
}
\catcode`\@=12 % @ signs are no longer letters
\newlength\listtextwidth

\catcode`\@=11 % @ signs are now treated as letters
\newlength{\@tabfninsert}
\newlength{\@tabfnwidth}
\newcommand{\tabfootnote}[2]{%
  \setlength{\@tabfninsert}{0.8em}
  \setlength{\@tabfnwidth}{\textwidth}
  \addtolength{\@tabfnwidth}{-\@tabfninsert}
  \addtolength{\@tabfnwidth}{-0.4em}
  \noindent\makebox[\@tabfninsert][r]{\footnotesize$^{#1}$\hfil}\hfill%
  \parbox[t]{\@tabfnwidth}{\footnotesize #2\hfill}}
\catcode`\@=12 % @ signs are no longer letters

\def\citeCTD{{\cite{%
nim:a279:290,*npps:b32:181,*nim:a338:254%
}}\xspace}
\def\citeCAL{{\cite{%
nim:a309:77,*nim:a309:101,*nim:a321:356,*nim:a336:23%
}}\xspace}

\renewcommand{\ZcoosysB}{%
The ZEUS coordinate system is a right-handed Cartesian system, with the $Z$
axis pointing in the proton beam direction, referred to as the ``forward
direction'', and the $X$ axis pointing towards the centre of HERA.
The coordinate origin is at the nominal interaction point.\xspace}

\begin{document}
%------------------------------------------------------------------------------
%       Title sheet
%------------------------------------------------------------------------------
\prepnum{DESY--09--142.rev}

\title{
Measurement of isolated photon production in deep inelastic $\mathbf{ep}$ scattering
}

\author{ZEUS Collaboration}
\draftversion{2.41}
\date{\today}

\abstract{
Isolated photon production in deep inelastic $ep$ scattering has been
measured with the ZEUS detector at HERA using an integrated luminosity
of $320\, \mathrm{pb}^{-1}$. Measurements were made  in the  isolated-photon transverse-energy and pseudorapidity ranges $4 < E_T^{\gamma} < 15\,\mathrm{GeV}$  and $-0.7 < \eta^{\gamma} < 0.9$ for  exchanged photon virtualities, $Q^2$, in the range $10 < Q^2 < 350\, \mathrm{GeV}^2$ and for invariant masses of the hadronic system $W_X> 5\,\mathrm{GeV}$.  Differential cross sections are presented for inclusive isolated photon 
production as functions of  $Q^2$, $x$, $E_T^{\gamma}$ and
$\eta^{\gamma}$.  Leading-logarithm parton-shower Monte Carlo
simulations  and  perturbative QCD predictions give a reasonable
description of the data over most of the kinematic range.
}

\makezeustitle

%------------------------------------------------------------------------------
%       Text
%------------------------------------------------------------------------------
\pagenumbering{arabic} 
\pagestyle{plain}

\begin{center}                                                                                     
{                      \Large  The ZEUS Collaboration              }                               
\end{center}                                                                                       
  S.~Chekanov,                                                                                     
  M.~Derrick,                                                                                      
  S.~Magill,                                                                                       
  B.~Musgrave,                                                                                     
  D.~Nicholass$^{   1}$,                                                                           
  \mbox{J.~Repond},                                                                                
  R.~Yoshida\\                                                                                     
 {\it Argonne National Laboratory, Argonne, Illinois 60439-4815, USA}~$^{n}$                       
\par \filbreak                                                                                     
  M.C.K.~Mattingly \\                                                                              
 {\it Andrews University, Berrien Springs, Michigan 49104-0380, USA}                               
\par \filbreak                                                                                     
  P.~Antonioli,                                                                                    
  G.~Bari,                                                                                         
  L.~Bellagamba,                                                                                   
  D.~Boscherini,                                                                                   
  A.~Bruni,                                                                                        
  G.~Bruni,                                                                                        
  F.~Cindolo,                                                                                      
  M.~Corradi,                                                                                      
\mbox{G.~Iacobucci},                                                                               
  A.~Margotti,                                                                                     
  R.~Nania,                                                                                        
  A.~Polini\\                                                                                      
  {\it INFN Bologna, Bologna, Italy}~$^{e}$
% ...
\newpage                                                                                           
\enlargethispage{5cm}                                                                              
$^{\    1}$ also affiliated with University College London,                                        
United Kingdom\\                                                                                   
% ...
$^{\    5}$ also Senior Alexander von Humboldt Research Fellow at Hamburg University,              
\mbox{Institute} of \mbox{Experimental} Physics, Hamburg, Germany\\                                           
% ...
%                                                                                                  
% \par         % if index listing & table fit to 1 page, put gap here                              
\newpage   % alternatively: go to newpage, if page is too small                                    
                                                           %                                       
% \institute_references_start    % do not touch or move this line !                                
                                                           %                                       
\begin{tabular}[h]{rp{14cm}}                                                                       
$^{a}$ &  supported by the Natural Sciences and Engineering Research Council of Canada (NSERC) \\  
$^{b}$ &  supported by the German Federal Ministry for Education and Research (BMBF), under        
          contract Nos. 05 HZ6PDA, 05 HZ6GUA, 05 HZ6VFA and 05 HZ4KHA\\                            
$^{c}$ &  supported in part by the MINERVA Gesellschaft f\"ur Forschung GmbH, the Israel Science   
          Foundation (grant No. 293/02-11.2) and the US-Israel Binational Science Foundation \\    
$^{d}$ &  supported by the Israel Science Foundation\\                                             
$^{e}$ &  supported by the Italian National Institute for Nuclear Physics (INFN) \\                
$^{f}$ &  supported by the Japanese Ministry of Education, Culture, Sports, Science and Technology 
          (MEXT) and its grants for Scientific Research\\                                          
$^{g}$ &  supported by the Korean Ministry of Education and Korea Science and Engineering          
          Foundation\\                                                                             
$^{h}$ &  supported by the Netherlands Foundation for Research on Matter (FOM)\\                   
$^{i}$ &  supported by the Polish State Committee for Scientific Research, project No.             
          DESY/256/2006 - 154/DES/2006/03\\                                                        
$^{j}$ &  partially supported by the German Federal Ministry for Education and Research (BMBF)\\   
$^{k}$ &  supported by RF Presidential grant N 1456.2008.2 for the leading                         
          scientific schools and by the Russian Ministry of Education and Science through its      
          grant for Scientific Research on High Energy Physics\\                                   
$^{l}$ &  supported by the Spanish Ministry of Education and Science through funds provided by     
          CICYT\\                                                                                  
$^{m}$ &  supported by the Science and Technology Facilities Council, UK\\                         
$^{n}$ &  supported by the US Department of Energy\\                                               
$^{o}$ &  supported by the US National Science Foundation. Any opinion,                            
findings and conclusions or recommendations expressed in this material                             
are those of the authors and do not necessarily reflect the views of the                           
National Science Foundation.\\                                                                     
$^{p}$ &  supported by the Polish Ministry of Science and Higher Education                         
as a scientific project (2009-2010)\\                                                              
$^{q}$ &  supported by FNRS and its associated funds (IISN and FRIA) and by an Inter-University    
          Attraction Poles Programme subsidised by the Belgian Federal Science Policy Office\\     
$^{r}$ &  supported by an FRGS grant from the Malaysian government\\                               
\end{tabular}                                                                                      
                                                           %                                       
% \institute_references_end     % do not touch or move this line !                                 
                                                           %                                       
%\end{document}                                                                                     

\section{Introduction}
\label{sec-int}

In the study of high-energy collisions involving hadrons, events in
which an isolated high-energy photon is observed provide a direct
probe of the underlying partonic process, since the emission of these
photons is unaffected by parton hadronisation. 

Isolated high-energy
photon production has been studied in a number of fixed-target and hadron-collider experiments \cite{zfp:c13:207,*zfp:c38:371,*pr:d48:5,*prl:73:2662,*prl:95:022003,*prl:84:2786,*plb:639:151}.  Previous ZEUS and H1 publications have also reported the production of isolated photons in
photoproduction~\cite{pl:b413:201,pl:b472:175,pl:b511:19,epj:c49:511,epj:c38:437}, in which the exchanged photon is quasi-real ($Q^2 \approx 0$), and deep inelastic scattering (DIS)~\cite{pl:b595:86,epj:c54:371}, in which \mbox{$Q^2 \approx \mathrm{GeV}^2$}.

Isolated photons are produced in DIS at lowest order in QCD as shown
in Fig.~\ref{fig1}. Photons produced by radiation from an incoming or
outgoing quark are called ``prompt''; an additional class of
high-energy photons comprises those radiated from the incoming or
outgoing lepton.  In this letter, results are presented from a new inclusive
measurement of isolated photon production in neutral current DIS.
The data provide a test of perturbative QCD in a kinematic region with two 
hard scales: $Q^2$, the exchanged photon virtuality,
and $E_T^{\gamma}$, the transverse energy of the emitted
photon. Compared to the previous ZEUS publication \cite{pl:b595:86},
the kinematic reach  extends to lower values of
$Q^2$ and to higher values of $E_T^{\gamma}$. The statistical
precision is also improved.

Leading-logarithm parton-shower Monte Carlo (MC) and perturbative QCD
predictions are compared to the measurements.  
The cross sections for isolated photon production in DIS 
have been calculated to order ${O} {(\alpha^3)}$ by
Gehrmann-De Ridder {\it et al.} (GGP)~\cite{np:b578:326,prl:96:132002,epj:c47:395}.  A calculation based on QED contributions to the parton
distributions has been made by Martin {\it et al.} (MRST)~\cite{epj:c39:155}.

\section{Experimental set-up}
\label{sec-exp}
The measurements are based on  a data sample corresponding to an integrated luminosity of $320\pm8\,\mathrm{pb}^{-1}$,  taken
between 2004 and 2007 with the ZEUS detector at HERA. The sample is a sum of $131\pm3  \,\mathrm{pb}^{-1}$ of $e^+p$ data and $189\pm 5 \,\mathrm{pb}^{-1}$ of $e^-p$ data\footnote{Hereafter `electron' refers to both
   electrons and positrons unless otherwise specified.} %taken at
with centre-of-mass energy $\sqrt{s}=318\,\mathrm{GeV }$.

A detailed description of the ZEUS detector can be found
elsewhere~\cite{zeus:1993:bluebook}. Charged particles were tracked in
the central tracking detector (CTD)~\citeCTD and a silicon micro vertex 
detector (MVD)~\cite{nim:a581:656} which operated in a
magnetic field of $1.43\,\Tesla$ provided by a thin superconducting
solenoid.  The high-resolution uranium--scintillator
calorimeter (CAL)~\citeCAL consisted of three parts: the forward (FCAL), the
barrel (BCAL) and the rear (RCAL) calorimeters. The BCAL covers the pseudorapidity range -0.74 to 1.01 as seen from the nominal interaction point.  The FCAL and RCAL extend the range to -3.5 to 4.0.  The smallest
subdivision of the CAL was called a cell. The barrel electromagnetic
calorimeter (BEMC) cells had a pointing geometry aimed at the nominal interaction point, with a cross section approximately $5\times20\,\mathrm{cm^2}$,
with the finer granularity in the $Z$-direction \ZcoosysfnB.  This fine granularity allows the use of shower-shape distributions to distinguish isolated photons from the
     products of neutral meson decays such as $\pi^0 \rightarrow \gamma \gamma$.

 A three-level trigger system was used to select events
online \cite{uproc:chep:1992:222,*nim:a580:1257} by requiring  well isolated electromagnetic deposits in the CAL. 

The luminosity was measured using the Bethe--Heitler reaction $ep \rightarrow e\gamma p$ by a luminosity detector which consisted of two independent systems: a lead--scintillator calorimeter \cite{desy-92-066,*zfp:c63:391,*acpp:b32:2025} and a magnetic spectrometer \cite{nim:a565:572}.

\section{Event selection and reconstruction}
\label{sec-selec}

Events were selected offline by requiring a scattered-electron candidate,
identified using a neural network~\cite{nim:a365:508,*nim:a391:360}.  The candidates were required to have a polar angle in the range
\mbox{$139.8^{\circ} < \theta_e < 171.9^{\circ}$} in order
to ensure that they were well measured in the RCAL. The impact point ($X$,$Y$)
of the candidate on the surface of the RCAL was required to lie outside the region ($\pm 15\,\mathrm{cm}$,$\pm 15\,\mathrm{cm}$) centred on (0,0) to ensure 
well understood acceptance. 
 The energy of the candidate, $E'_e$, was required to be larger than
$10\,\mathrm{GeV}$. The kinematic quantities $Q^2$ and $x$ were reconstructed from the scattered electron by means of the relationships $Q^2=-(k-k')^2$ and \mbox{$x=Q^2/(2P\cdot(k-k'))$} where $k$ ($k'$) is the four-momentum of the incoming (outgoing) lepton  and $P$ is the four-momentum of the incoming proton.  The kinematic region $10 <Q^2< 350\,\mathrm{GeV^2}$ was selected. 

To reduce backgrounds from non-$ep$ collisions, events
were required to have a reconstructed vertex position,
$Z_{\mathrm{vtx}}$, within the range $|Z_{\mathrm{vtx}}|<
40\,\mathrm{cm}$ and to have $35 < \delta < 65\,\mathrm{GeV}$, where
$\delta = \sum \limits_i E_i(1-\cos \theta_i)$; $E_i$ is the energy of
the $i$th CAL cell, $\theta_i$ is its polar angle and the sum runs
over all cells~\cite{pl:b303:183}.  At least one reconstructed track, well separated from
the electron, was required, ensuring some hadronic activity which suppressed
deeply virtual Compton scattering (DVCS) \cite{pl:b517:47,*jhep:0905:108} to a negligible level.

Photon candidates were identified as CAL energy-flow objects (EFOs)
\cite{epj:c1:81,*epj:c6:43} for which at least $90\%$ of the
reconstructed energy was measured in the BEMC.  EFOs with
wider electromagnetic showers than are typical of a single photon were
accepted to allow evaluation of backgrounds.  The reconstructed transverse energy of the EFO, $E_T^{\gamma}$, was required to lie
within the range \mbox{$4 <E_T^{\gamma}<15\,\mathrm{GeV}$} and the
pseudorapidity, $\eta^{\gamma}$, had to satisfy $-0.7 < \eta^{\gamma}
< 0.9$. The upper limit on the reconstructed transverse energy was
selected to ensure that the shower shapes from background and signal
remained distinguishable. 

 To reduce the background from photons and neutral mesons within jets, the EFO was required to be isolated from reconstructed tracks and hadronic activity.  Isolation from tracks was initially achieved by demanding
$\Delta R>0.2$, where $\Delta R = \sqrt{(\Delta \phi)^2 +
  (\Delta\eta)^2}$ is the distance to the 
nearest reconstructed track with momentum greater than
$250\,\mathrm{MeV}$  in the $\eta \-- \phi$ plane, where $\phi$ is the azimuthal angle.  Jet reconstruction was performed on all EFOs in the event, including
the electron and photon candidates, using the $k_T$ cluster algorithm
\cite{np:b406:187} in the longitudinally invariant inclusive mode
\cite{pr:d48:3160} with $R$ parameter set to 1.0.  Further isolation was imposed by requiring that
the photon-candidate EFO possessed at least $90 \%$ of
the total energy of the jet of which it formed a part.

Each event was required to contain both an electron and a photon
candidate. The invariant mass of the hadronic system, $W_X$, is then defined by
  $W^2_X= (P + k -k' - p_{\gamma})^2 $ , where $p_{\gamma}$ is the four-vector of the outgoing photon.
 A total of 15699 events were selected; at this stage the
sample was dominated by background events.  The largest source of
background was neutral current (NC) DIS events where a genuine electron
candidate was found in the RCAL and neutral mesons, such as $\pi^0$ and
$\eta$, decaying to photons, produced a photon-candidate EFO in the BEMC.

\section{Theory}
\label{sec:theory}
\begin{figure}[p]
\vfill
\begin{center}
\epsfig{file=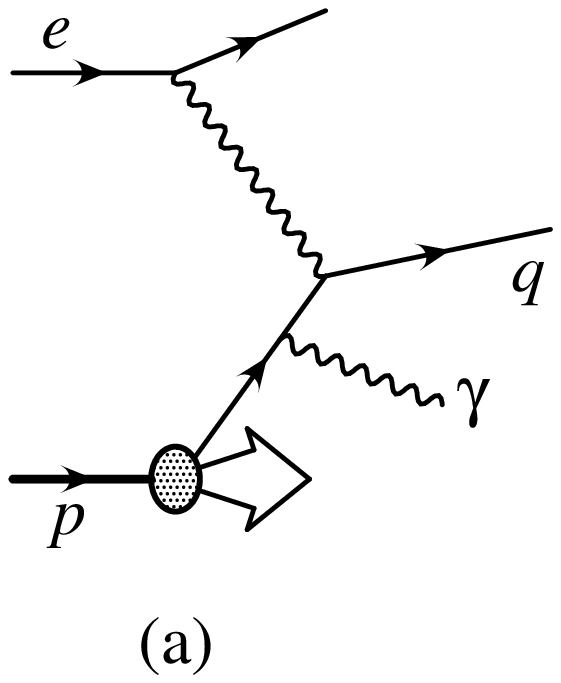,width=6cm}
\epsfig{file=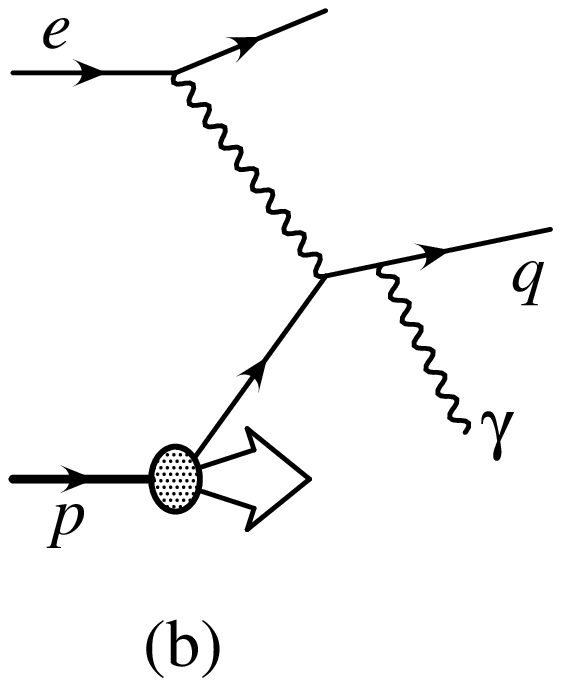,width=6cm}\\
\vspace{1.5cm}
\epsfig{file=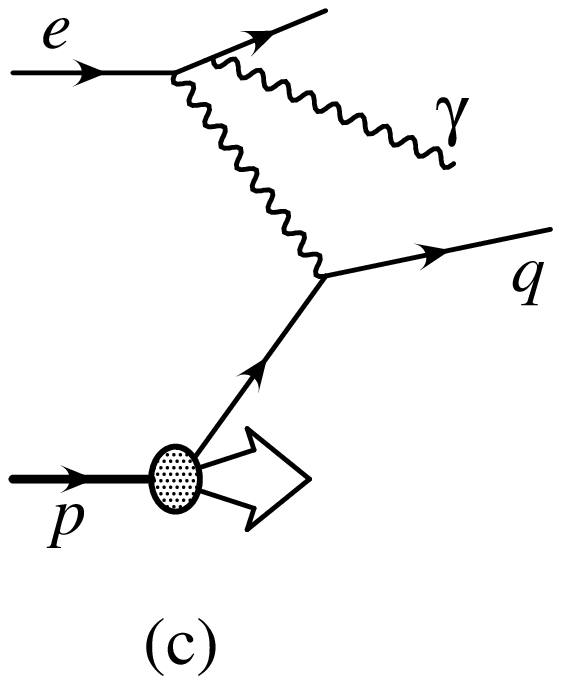,width=6cm}
\epsfig{file=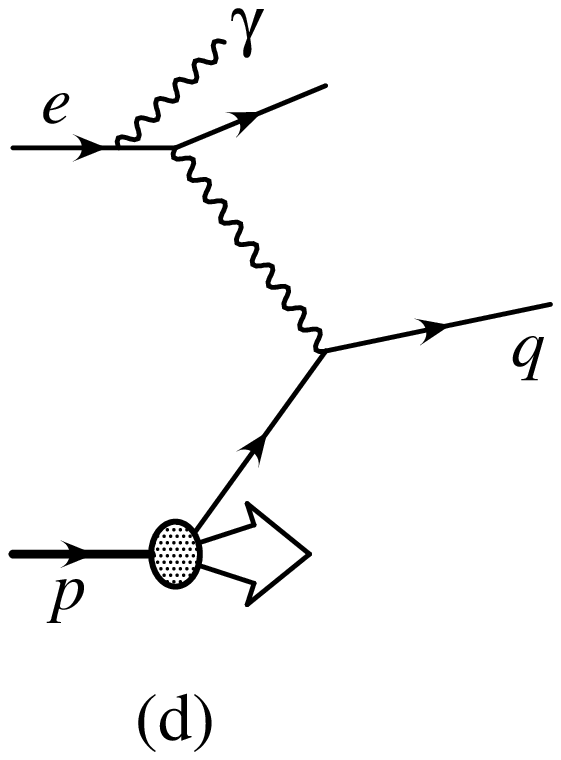,width=6cm}
\vspace{1.5cm}
\end{center}
\caption{\small Lowest-order tree-level diagrams for isolated photon 
production in 
$ep$ scattering. %Vertex corrections enter at the same order.
\label{fig1}}
\vfill
\end{figure}

Two theoretical predictions are compared to the measurements presented
in this paper. In the
approach of GGP~\cite{prl:96:132002}, the contributions to the
scattering cross section for \mbox{$ep \rightarrow e\gamma X$} are calculated at order $\alpha^3$ in the electromagnetic coupling. One of
these contributions comes from the radiation of a photon from the
quark line (called QQ photons; Fig.~\ref{fig1}a,b) and a second from the radiation from
the lepton line (called LL photons; Fig.~\ref{fig1}c,d). In addition to QQ and LL photons,
the interference term between photon emission from the lepton and quark lines,
called LQ photons by GGP, is evaluated.  For the
kinematic region considered here, where the outgoing photon is well
separated from both outgoing electron and quark, the interference term
gives only a 3\% effect on the cross section. This effect is further reduced
to $\approx{1\%}$ when $e^+p$ and $e^-p$ data are combined as the LQ term
changes sign when $e^-$ is replaced by $e^+$.  The QQ contribution includes both wide-angle photon emission and the leading $q\to q \gamma$ fragmentation term. GGP have chosen to use CTEQ6L leading-order parton distribution functions~\cite{jhep:0207:012}. The factorisation scales used are $Q$ for QQ events and max($Q$, $\mu_{F,min}$) for LL events where $\mu_{F,min}= 1\, \mathrm{GeV}$.  Parton-to-hadron corrections were not made, in view of technical issues in relating $2\to 2$ and $2 \to 3$  topologies, following the advice of the GGP authors. We note that others have taken a different view~\cite{epj:c54:371}. A na\"{i}ve study indicated the likely effect to be a reduction after hadronisation in predicted inclusive cross-sections of order $15\%$.

In the approach of MRST~\cite{epj:c39:155,priv:stirling:2009}, a partonic photon component of the proton, $\gamma_p$, is introduced as a consequence of
including QED corrections in the parton distribution
functions. This leads to $ep$ interactions taking place via QED
Compton scattering,  $\gamma_p  e \to \gamma e$. A measurement of the
isolated high-energy photon production cross section therefore
provides a constraint on the photon density in the proton.  The model
includes the collinearly divergent LL contribution, which is enhanced relative to that of GGP by the DGLAP resummation due to the inclusion
 of QED Compton scattering. The QQ component is not included in 
the MRST model, in which the transverse momentum of  the scattered electron is 
expected to balance approximately that of the isolated photon. In the analysis
 presented here, such a constraint was not imposed.  The theoretical
 uncertainties in the models have been estimated by varying the
 factorisation scales by a factor two.

Since the MRST cross sections include the LL contribution of GGP
to a good approximation, but exclude the QQ, an improved prediction
can be constructed by summing the MRST cross section and the QQ
cross section from GGP~\cite{priv:gehrmann:2009,priv:stirling:2009}.  The theory uncertainties are of the same order as those of the
individual QQ and LL components.

\section{Monte Carlo event simulation}
\label{sec-mc}
The MC program {\sc Pythia} 6.416 \cite{jhep:0605:026} was used to
simulate prompt-photon emission for the study of the event-reconstruction
efficiency. In {\sc Pythia}, this process is simulated as a DIS process
with additional photon radiation from the quark line to account for
QQ photons.  Radiation from the lepton is not simulated in this {\sc Pythia}
sample.

The LL photons radiated at large angles from the incoming or outgoing
electron were simulated using the generator {\sc Djangoh} 6
\cite{cpc:81:381}, an interface to the MC program {\sc Heracles} 4.6.6
\cite{cpc:69:155}; higher-order QCD effects were included using the
colour dipole model of {\sc Ariadne} 4.12\cite{cpc:71:15}. Hadronisation of the partonic final state was
performed by {\sc Jetset} 7.4 \cite{cpc:39:347}. The small LQ contribution was
neglected.

 The NC DIS background was simulated using {\sc Djangoh} 6, within the same 
framework as the LL events. This provided a realistic spectrum of mesons and overlapping clusters  with well modelled kinematic distributions and hence was preferred to single-particle MC samples for backgrounds, such as were used in
the previous ZEUS publication~\cite{pl:b595:86}.

The MC samples described above contained only events in which $W_X$ was larger than $5\,\mathrm{GeV}$. 
Isolated photons can also be produced at values of $W_X$ less than $5\,\mathrm{GeV}$  in  `elastic' and `quasi-elastic' processes ($ep\rightarrow ep\gamma$) such as DVCS and Bethe--Heitler photon production. Such 
events were simulated using the {\sc GenDVCS}\cite{misc:gendvcs} and {\sc Grape-Compton}~\cite{cpc:136:126} 
generators. The contribution of these elastic processes was negligible after
the selections described in Section \ref{sec-selec}.

The generated MC events were passed through the ZEUS detector and
trigger simulation programs based on {\sc Geant} 3.21
\cite{tech:cern-dd-ee-84-1}. They were reconstructed and analysed by
the same programs as the data.  In addition to the full-event
simulations, MC samples of single particles (photons and neutral
mesons) were generated and used to study the MC description of
electromagnetic showering in the BEMC.

\section{Extraction of the photon signal}

The event sample selected according to the criteria in Section
\ref{sec-selec} was dominated by background; thus the photon signal was
extracted statistically following the approach used in previous ZEUS
analyses \cite{pl:b413:201,pl:b472:175,pl:b511:19,pl:b595:86}.

The photon signal was extracted from the background using BEMC
energy-cluster shapes. Two shape variables were considered:
\begin{itemize}
\item the variable  
$\langle \delta Z\rangle= \frac{\sum \limits_i
E_i|Z_i-Z_{\mathrm{cluster}}| }{w_{\mathrm{cell}}\sum \limits_i E_i}$,
where $Z_{i}$ is the $Z$ position of the centre of the $i$th cell,
$Z_{\mathrm{cluster}}$ is the centroid of the EFO cluster,
$w_{\mathrm{cell}}$ is the width of the cell in the $Z$ direction,
$E_i$ is the energy recorded in the cell and the sum runs over all
BEMC cells in the EFO;
\item the ratio  $f_{\mathrm{max}}$ of the highest energy deposited in 
any one BEMC cell in the EFO to the total EFO BEMC energy.
\end{itemize}

\begin{figure}[p]

\vfill
\begin{center}
\epsfig{file=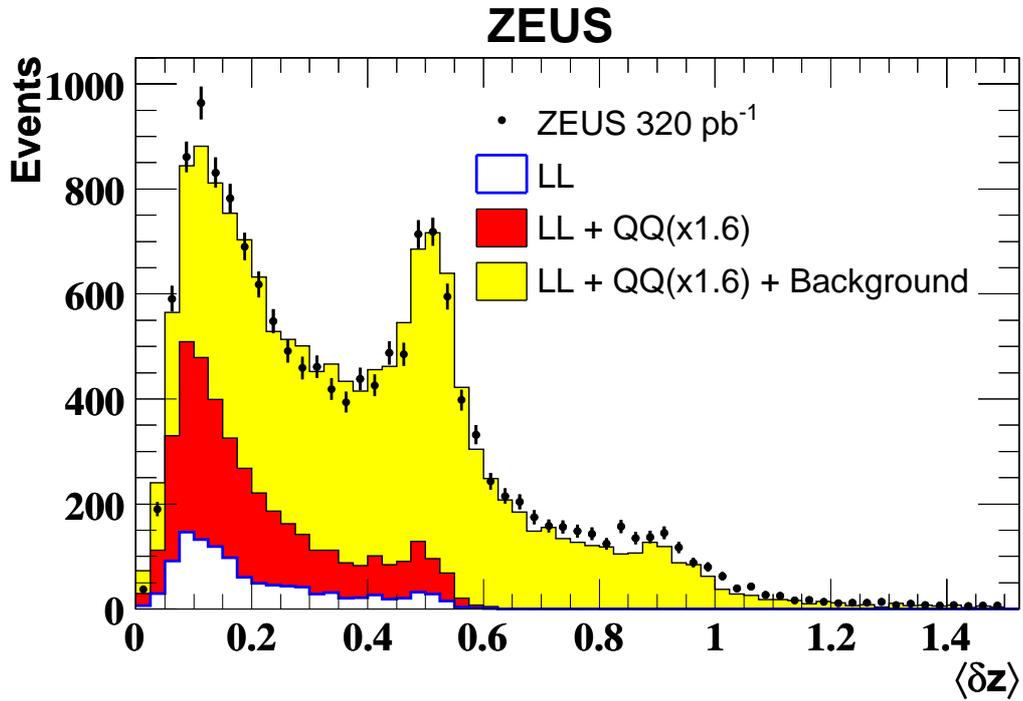,width=14cm}
\epsfig{file=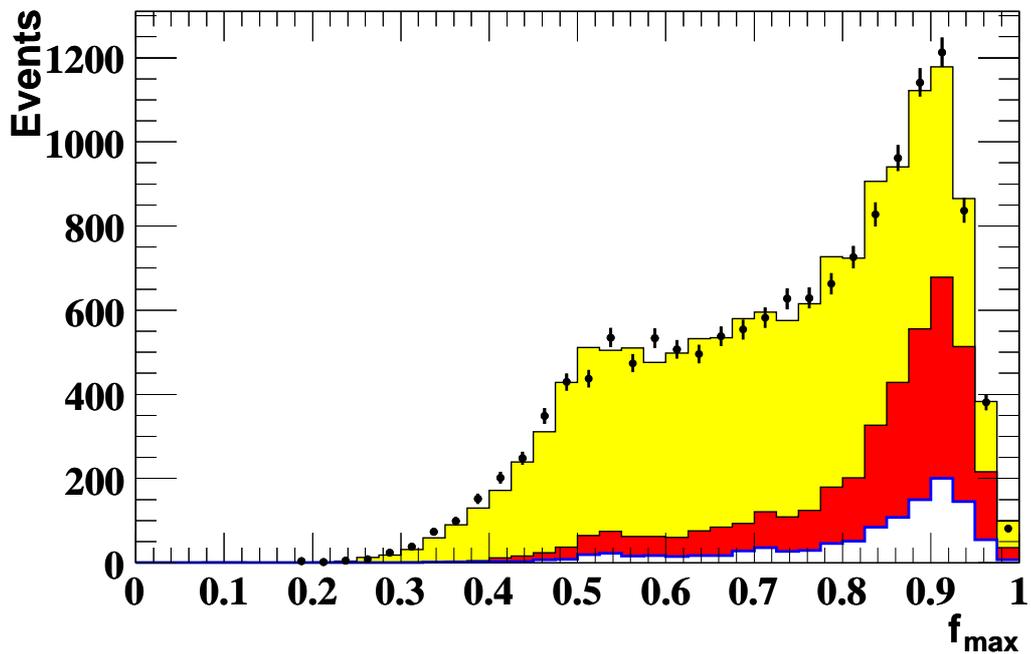,width=14cm}

\end{center}

\caption{\small 
 Distributions of $\langle \delta Z \rangle$ and
  $f_{\mathrm{max}}$. The error bars represent the statistical uncertainties.
 The light shaded histogram shows a fit to the data of three components
with fixed shapes as described in the text.
%      background from the {\sc Ariadne}-hadronic MC, with QQ photons from
 %     {\sc Pythia} and LL photons from {\sc Djangoh6}.  
The dark shaded histogram represents the QQ  component of the fit, and the white
      histogram the LL component.
% The solid curve is from a fit to the data
%of {\sc{Ariadne}}-hadronic MC and QQ photons from {\sc{Pythia}} (dark area). The 
%white area shows the predicted number of LL
%photons.  
The $f_{\mathrm{max}}$ distribution is shown after requiring
$\langle \delta Z \rangle < 0.8$.}
\label{fig:showers}
\vfill
\end{figure}

The distributions of $\langle \delta Z \rangle$ and $f_{\mathrm{max}}$
(after the requirement  $\langle \delta Z \rangle < 0.8$)
in the data and the MC are shown in Fig.\,\ref{fig:showers}. The MC LL and QQ
distributions have been corrected in each two-dimensional ($\eta$, $E_{T}$) bin using factors derived from the difference between simulated and real DIS electron data.  The $\langle \delta Z \rangle$ distribution exhibits a double-peaked
structure with the first peak at $\approx 0.1$, associated with the
signal, and a second peak at $\approx 0.5$, dominated by the $\pi^0
\rightarrow
\gamma \gamma$ background. The $f_{\mathrm{max}}$ distribution shows a single peak at
$\approx 0.9$ corresponding to the photon signal, and has a shoulder extending down to $\approx 0.5$, which is dominated by the hadronic background.

The number of isolated-photon events contributing to
Fig.\,\ref{fig:showers} and in each cross-section bin was determined by
a $\chi^2$ fit to the $\langle \delta Z \rangle$ distribution in the
range $0<\langle \delta Z \rangle < 0.8$ using the LL and QQ signal
and background MC distributions as described
in Section~\ref{sec-mc}.  By treating the LL and QQ photons separately, one automatically takes account of their differing hadronic activity, (resulting in significantly different acceptances) and their differing ($\eta$, $E_{T}$) distributions, (resulting in different bin migrations due to finite measuring precision).

In performing the fit, the LL contribution was kept constant at its MC-predicted value and the other components
were varied.  Of the 15699 events selected, $4164\pm 168$ correspond to the extracted signal
(LL and QQ). The scale factor resulting from the global fit for the QQ photons
 in Fig.\,\ref{fig:showers} was 1.6; this factor was used 
for all the plots comparing MC to data. The fitted global scale factor for the hadronic background was 1.0. The signal fraction in the
cross-section bins varied from $21\% $ to $62\% $. In all cross-section 
bins, the $\chi^2/\mathrm{n.d.f.}$ of the fits was 2.1 or
smaller.

For a given observable $Y$, the production cross section was determined
using:

\begin{equation}
\frac{d\sigma}{dY} = \frac{\mathcal{A}_{\mathrm{QQ}} \cdot N(\gamma_{\mathrm{QQ}})}{  \mathcal{L} \cdot \Delta Y} + \frac{d\sigma^{\mathrm{MC}}_{\mathrm{LL}}}{dY} \nonumber ,
\end{equation}

where $N(\gamma_{\mathrm{QQ}})$ is the number of QQ photons extracted from the
fit, $\Delta Y$ is the bin width, $\mathcal{L}$ is the total
integrated luminosity, $\sigma^{\mathrm{MC}}_{\mathrm{LL}}$ is the
predicted cross section for LL photons from {\sc Djangoh}, and
$\mathcal{A}_{\mathrm{QQ}}$ is the acceptance correction for QQ
photons.  The value of $\mathcal{A}_{\mathrm{QQ}}$ was calculated
using Monte Carlo  from the ratio of
the number of events generated to those reconstructed in a given bin.  It varied between 1.2 and 
1.7 from bin to bin.

The fits employed in this analysis were performed using  $\langle \delta Z \rangle$
because of the larger difference in shape between signal and background for this quantity.
Fits in terms of  the $f_\mathrm{max}$ distributions were 
performed as a cross-check and gave similar results.  As a further cross-check, an algorithm from
the previous ZEUS publication \cite{pl:b595:86}, which selects wider
electromagnetic clusters as photon candidates, was used. This proved to be more sensitive
to the modelling of calorimeter backgrounds.  In every case 
where a satisfactory fit was obtained, good agreement with the principal method was found. The corrections to
the MC photon-signal energy-cluster shapes gave changes to the results within the 
statistical uncertainties and were not further considered\cite{forrest:phd:2009}.

\section{Systematic uncertainties}

The following sources of systematic uncertainty were investigated \cite{forrest:phd:2009}:

\begin{itemize}
\item the energy scale of the electromagnetic calorimeter (EMC)  
was varied by its known scale uncertainty of $\pm 2\%$ causing
variations in the measured cross sections of typically less than $\pm
2\%$;
\item  the dependence on the modelling of the hadronic background by {\sc Ariadne}
was investigated by varying the upper limit for the  $\langle \delta Z
\rangle$ fit in the range $0.6\rnge1.0$, giving variations that were
typically $\pm5\%$ but up to +12\% and -14\% in the most forward $\eta^\gamma$ and highest-$x$ bins respectively.
\end{itemize}

The following sources of systematic uncertainty were also investigated
and found to be negligible compared to the statistical uncertainty\cite{forrest:phd:2009}:

\begin{itemize}
\item variation of the EMC energy-fraction cut for the photon
  candidate EFO by $\pm 5\%$;

\item variation of the $Z_{\mathrm{vtx}}$ cut by $\pm 5\,\mathrm{cm}$;

\item variation of the upper and lower cuts on $\delta$ by $\pm3\,\mathrm{GeV}$;
\item variation of the $\Delta R$ cut used for track isolation by $\pm 0.1$;
\item variation of the track-momentum cut used in calculating track isolation by $\pm 100\,\mathrm{MeV}$;
\item variation of the LL-signal component by $\pm 5\%$.
\end{itemize}

All the uncertainties listed above were added in quadrature to give 
separate positive and negative systematic uncertainties in each bin.  The uncertainty of $2.6\% $ on the luminosity measurement was not included in the differential cross sections but included in the
integrated cross sections.

\section{Results}

\label{sec:results}

The cross section for inclusive isolated photon production, $ep
\rightarrow e \gamma X$, was measured in the kinematic
region defined by: $10 < Q^2< 350 \, \mathrm{GeV}^2$, $W_X >5\,\mathrm{GeV}$,
$E'_e > 10\,\mathrm{GeV}$, $139.8^{\circ}< \theta_e < 171.8^{\circ}$,
$-0.7< \eta^\gamma < 0.9$ and $4 < E_T^{\gamma}< 15\,\mathrm{GeV} $,
with isolation such that at least $90 \%$ of the energy of the jet
containing the photon belongs to the photon, where jets were formed
according to the $k_T$ algorithm with $R$ parameter set 1.0. The measured integrated
cross section is
\begin{equation}
19.4 \pm 0.7\,(\mathrm{stat.}) ^{+1.2}_{-1.0}\,(\mathrm{syst.})\,\mathrm{pb} , \nonumber
\end{equation}   
with an extracted contribution for QQ of
\begin{equation}
12.2 \pm 0.7\,(\mathrm{stat.}) ^{+1.2}_{-1.0}\,(\mathrm{syst.})\,\mathrm{pb}. \nonumber
\end{equation} 

\begin{figure}[p]
\vfill
\begin{center}
\epsfig{file=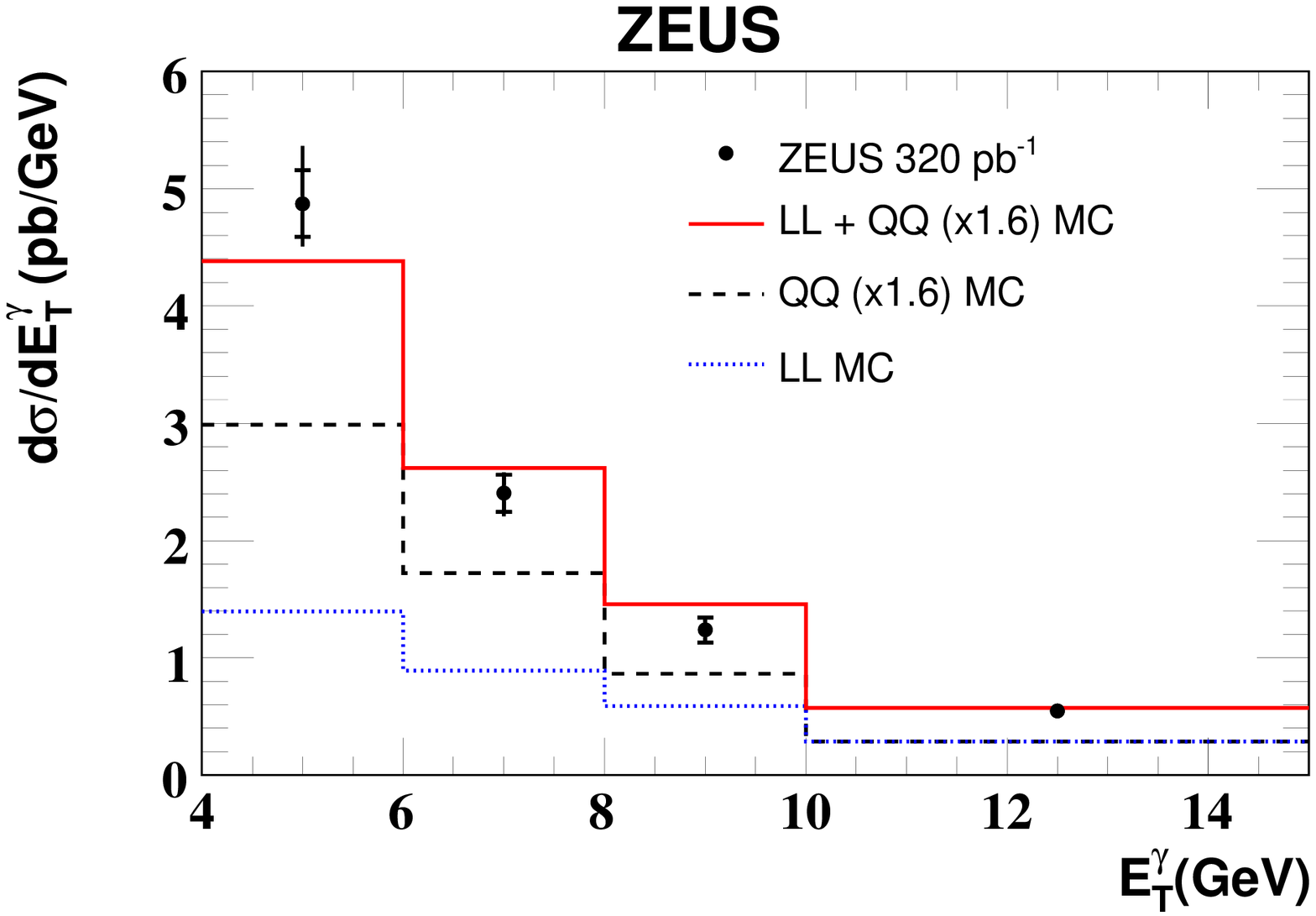,width=7.5cm}\\
\epsfig{file=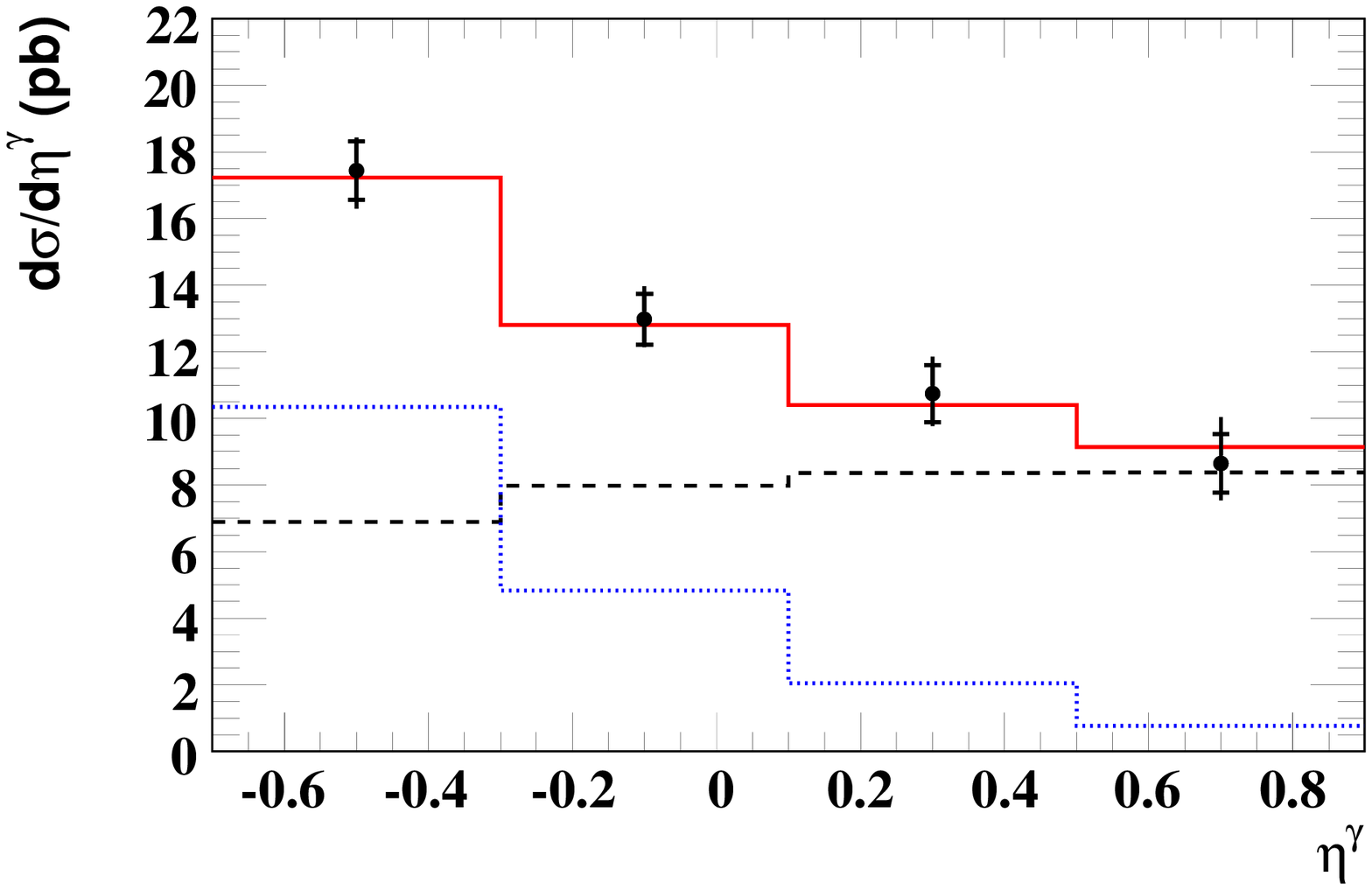,width=7.5cm}\\
\epsfig{file=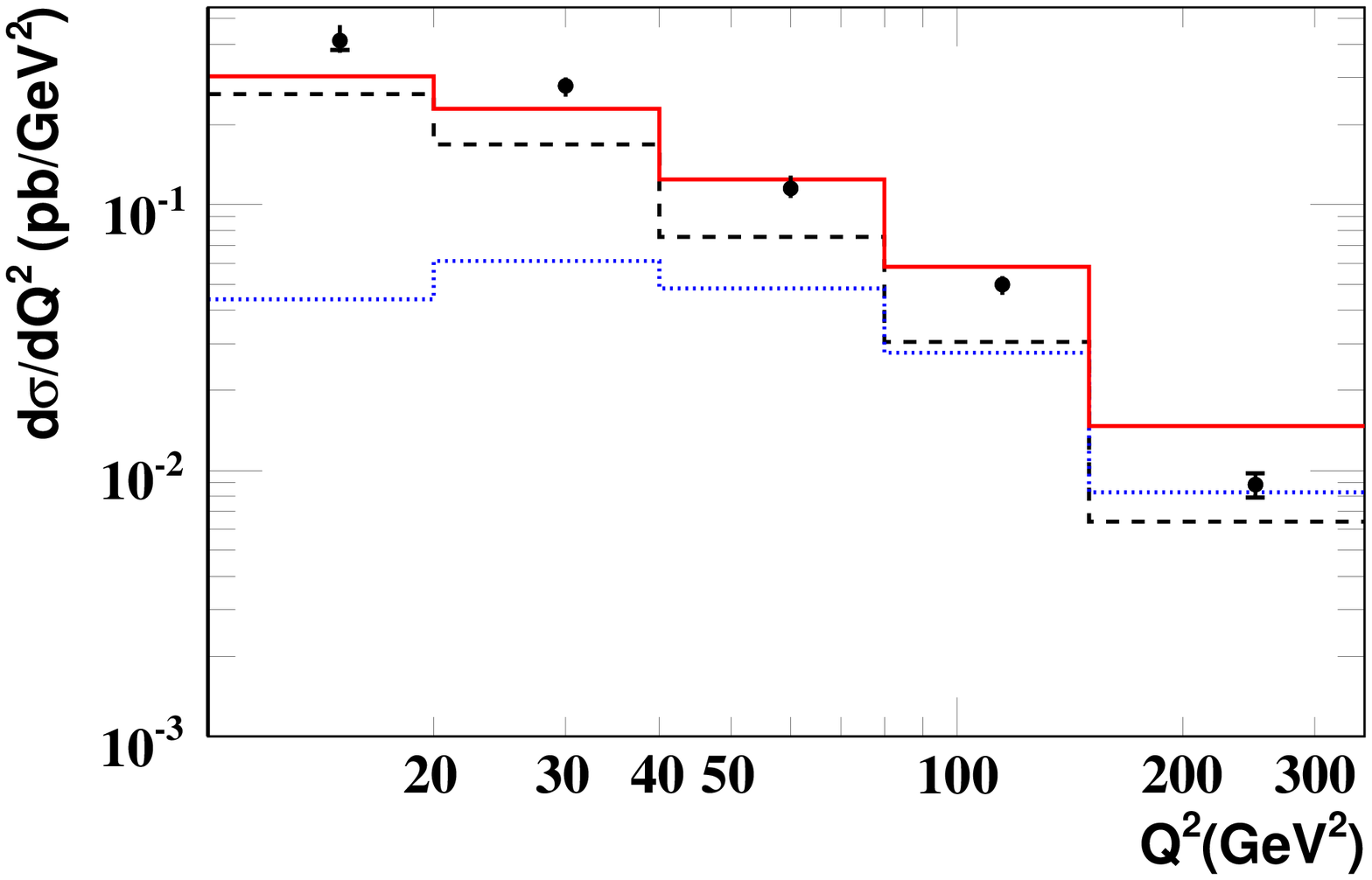,width=7.5cm}\\
\epsfig{file=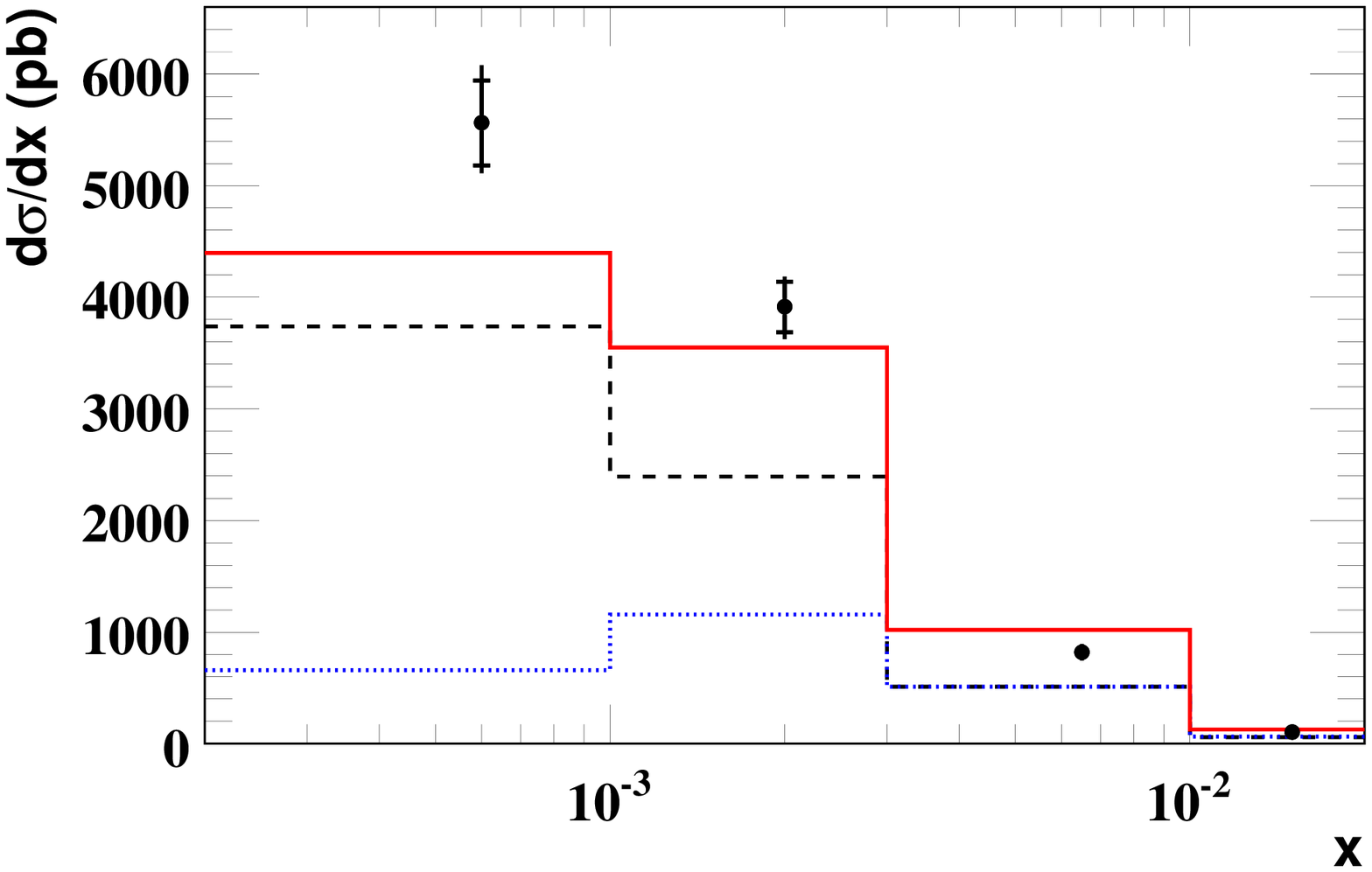,width=7.5cm}
\end{center}
{\vspace{-19.75cm} \bf\hspace{7.6cm}\ \ \ \ \ \ \ \ \ \ \ \ \ \ \ \ \ \ (a) \vspace{19.75cm}}\\

{\vspace{-15.9cm} \bf\hspace{7.6cm}\ \ \ \ \ \ \ \ \ \ \ \ \ \ \ \ \ \ (b) \vspace{15.9cm}}\\

{\vspace{-11.85cm} \bf\hspace{7.6cm}\ \ \ \ \ \ \ \ \ \ \ \ \ \ \ \ \ \ (c) \vspace{11.85cm}}

{\vspace{-7.35cm} \bf\hspace{7.6cm}\ \ \ \ \ \ \ \ \ \ \ \ \ \ \ \ \ \ (d) \vspace{3.25cm}}

\caption{\small 
Isolated photon differential cross sections in (a) $E_{T}^{\gamma}$, (b) $\eta^{\gamma}$, (c) $Q^{2}$ and (d) $x$.
The inner and outer error bars show, respectively, the statistical uncertainty and
the  statistical and systematic uncertainties 
added in quadrature. The solid histograms are the Monte Carlo
predictions from the sum of
QQ photons from {\sc Pythia} normalised by a factor 1.6 plus {\sc Djangoh} LL photons. The dashed (dotted)
lines show the QQ (LL) contributions.
\label{fig:xsec1}}
\vfill
\end{figure}

The differential cross sections as functions of $E_T^{\gamma}$,
$\eta^{\gamma}$, $Q^2$ and $x$  are shown in Fig.~\ref{fig:xsec1} and
given in Tables
\ref{tab:dsdet}--\ref{tab:dsdx}.  It can be seen that the cross section decreases with increasing $E_T^{\gamma}$, $\eta^{\gamma}$, $Q^2$ and $x$.  The predictions for the sum of the expected LL
contribution from {\sc Djangoh} and a factor of approximately 1.6
times the expected QQ contribution from {\sc Pythia} agree well with
the measurements, except for some differences at the lowest $Q^2$ (and
correspondingly lowest $x$).

\begin{figure}[p]
\vfill
\begin{center}
\epsfig{file=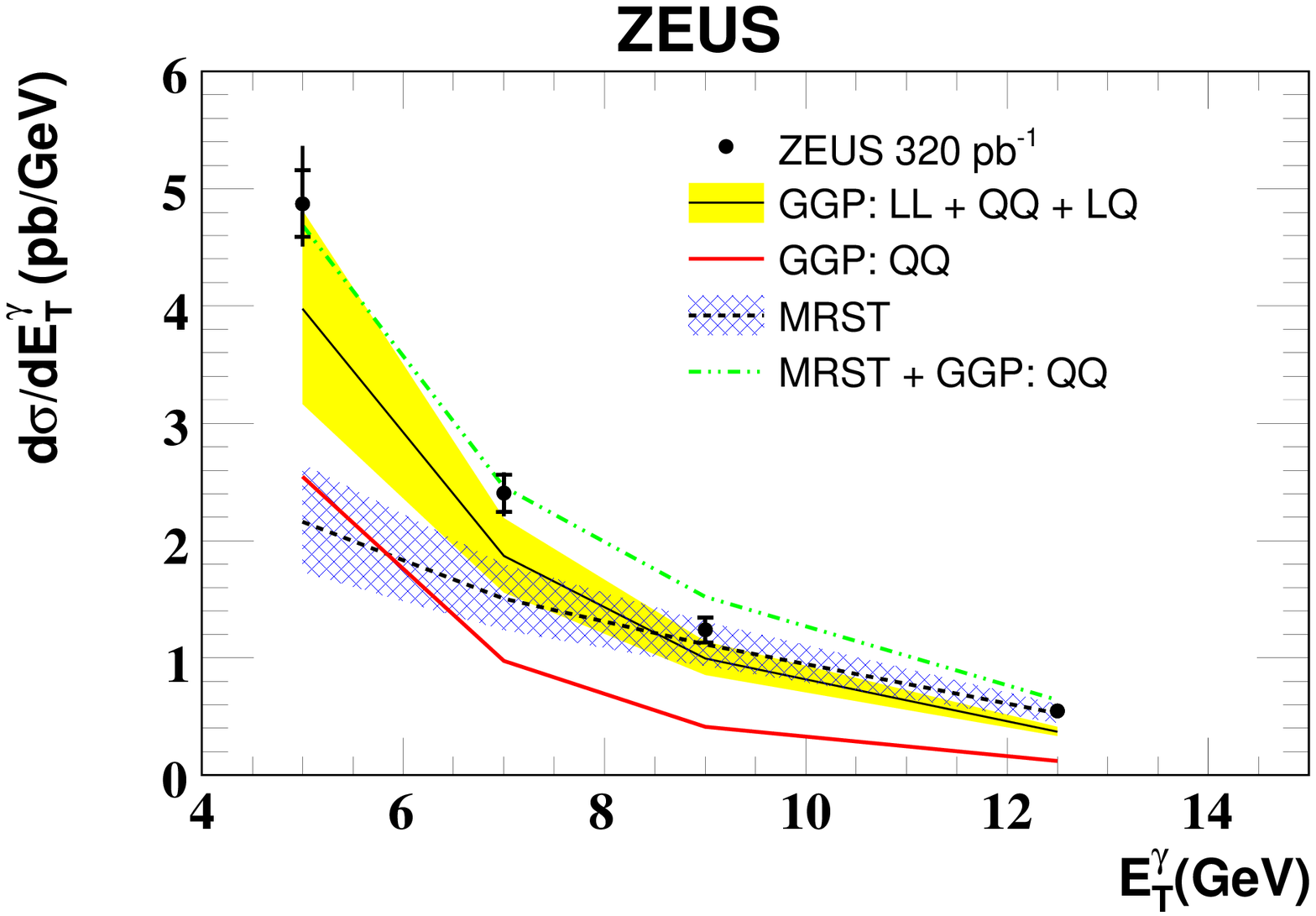,width=7.5cm}\\
\epsfig{file=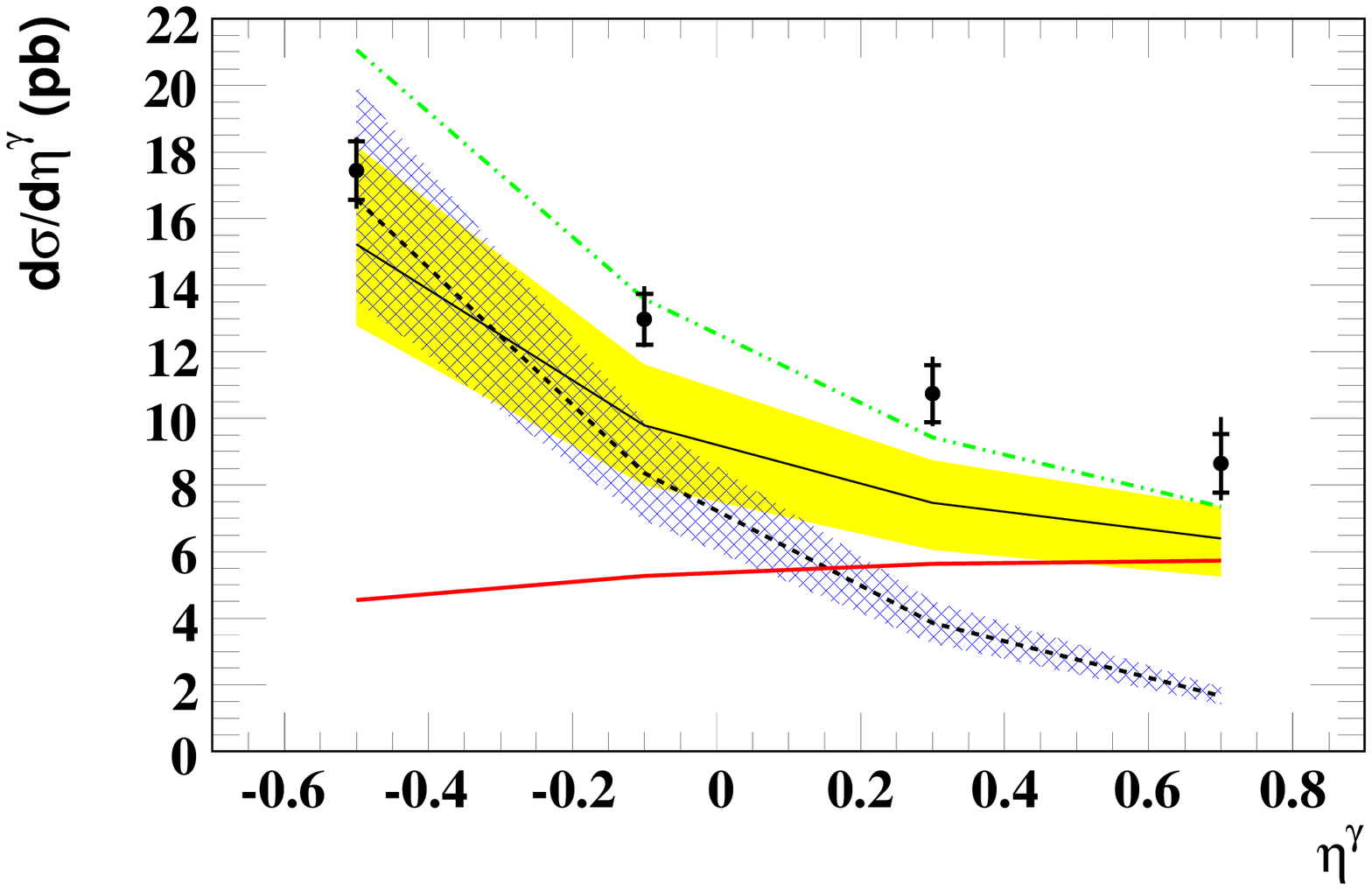,width=7.5cm}\\
\epsfig{file=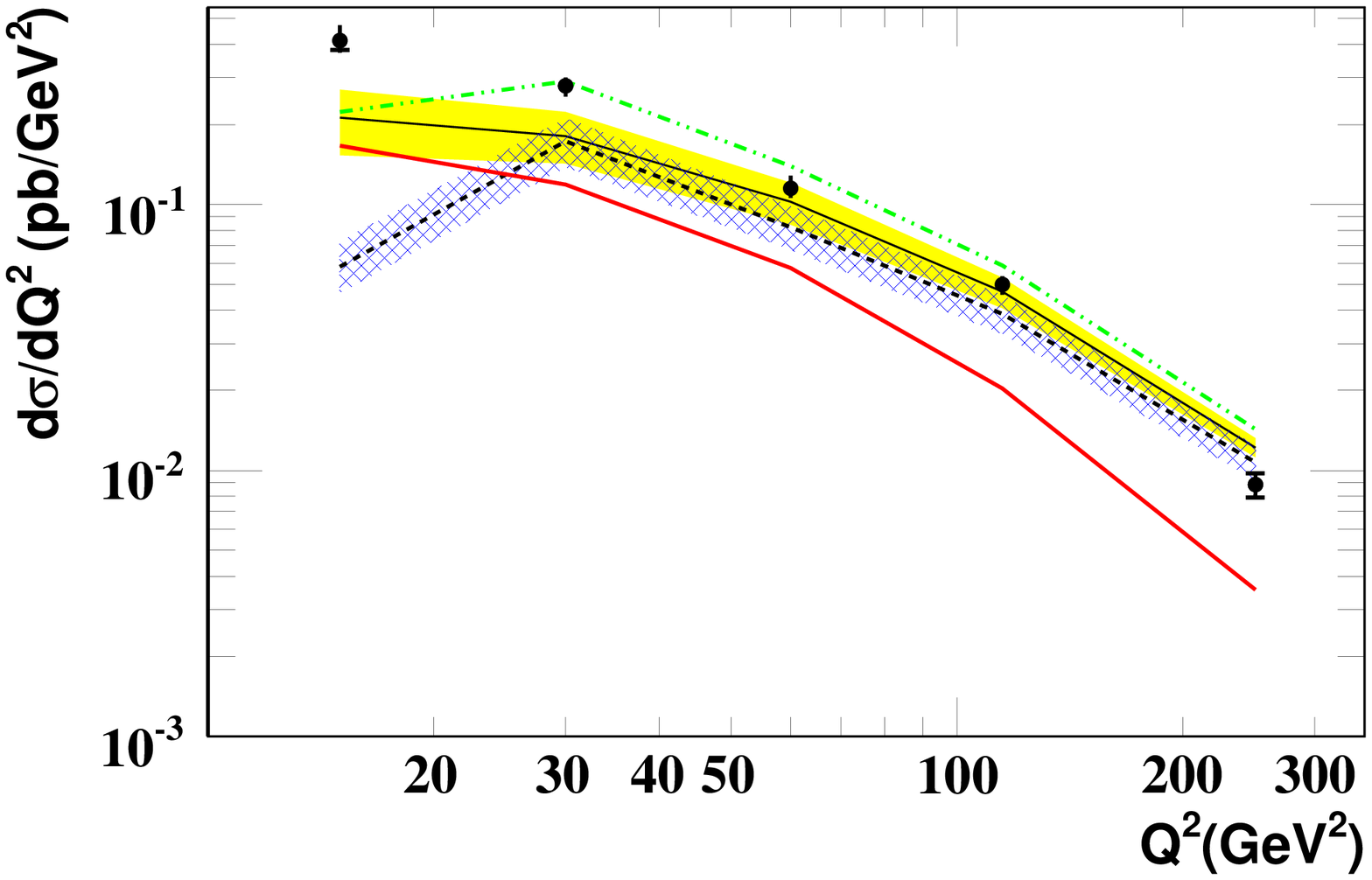,width=7.5cm}\\
\epsfig{file=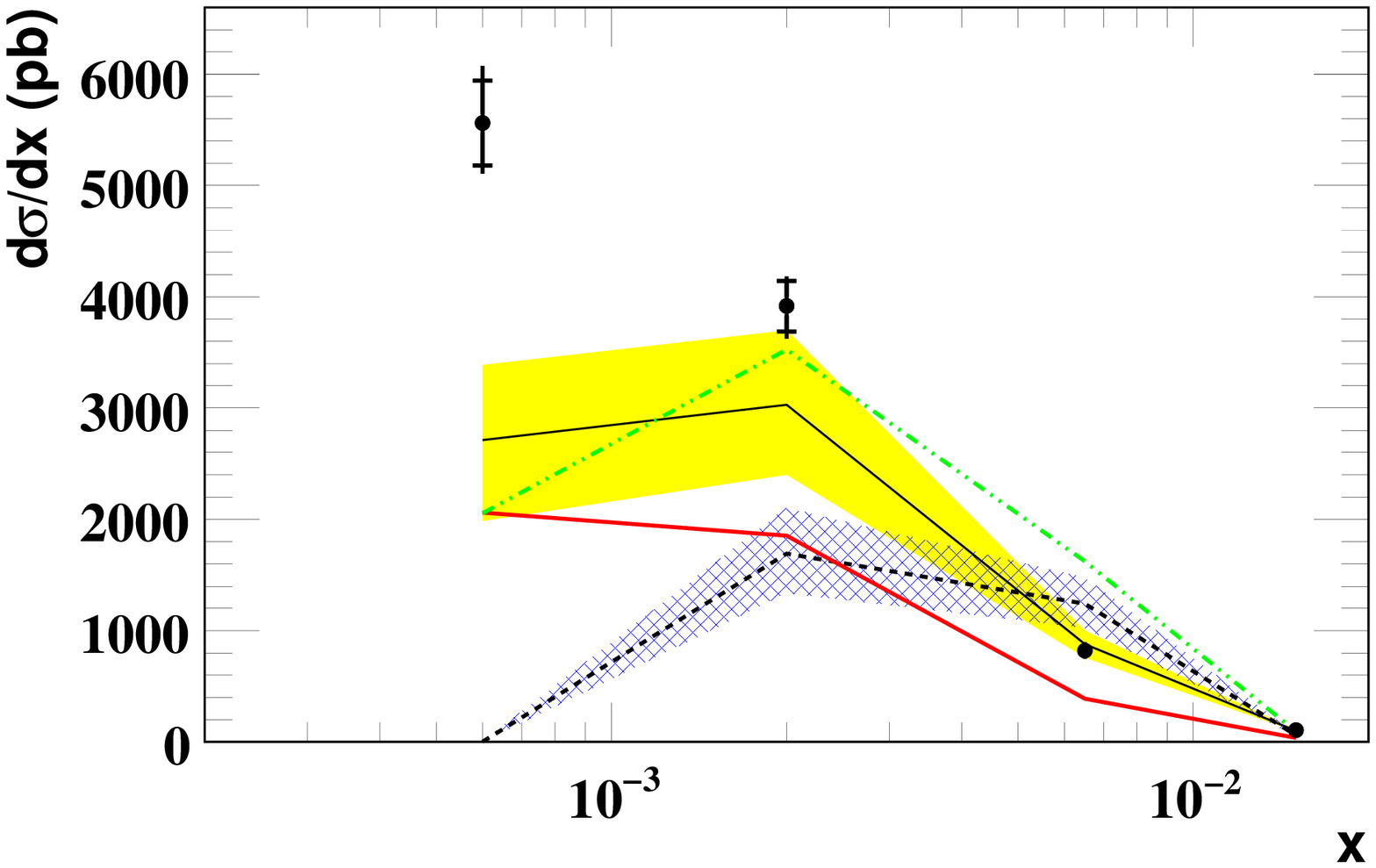,width=7.5cm}
\end{center}
{\vspace{-19.95cm} \bf\hspace{7.6cm}\ \ \ \ \ \ \ \ \ \ \ \ \ \ \ \ \ \ (a) \vspace{19.95cm}}\\

{\vspace{-16.0cm} \bf\hspace{7.6cm}\ \ \ \ \ \ \ \ \ \ \ \ \ \ \ \ \ \ (b) \vspace{16.0cm}}\\

{\vspace{-11.95cm} \bf\hspace{7.6cm}\ \ \ \ \ \ \ \ \ \ \ \ \ \ \ \ \ \ (c) \vspace{11.95cm}}

{\vspace{-7.45cm} \bf\hspace{7.6cm}\ \ \ \ \ \ \ \ \ \ \ \ \ \ \ \ \ \ (d) \vspace{3.35cm}}
\caption{\small 
Data points as Fig.~\ref{fig:xsec1}.
Theoretical predictions from Gehrmann-De Ridder et al.\ and Martin et al.\ are
shown with their associated uncertainties indicated by the shaded band
and the  hatched bands respectively.  The dash-dotted line illustrates
the combination MRST plus GGP: QQ.
\label{fig:xsec3}}
\vfill
\end{figure}

The theoretical predictions described in Section~\ref{sec:theory} are
compared to the measurements in Fig.~\ref{fig:xsec3}. The predictions from GGP describe
the shape of the $E_T^{\gamma}$ and $\eta^{\gamma}$ distributions
well, but their central value typically lies $20 \%$
below the measured cross sections.  The calculations fail to reproduce
the shape in $Q^2$; a similar observation was made by H1\cite{epj:c54:371}.  As with the MC comparison, the measured cross section is larger than
the theoretical prediction; this is also reflected in an excess of data
over theory at low $x$.

The MRST predictions mostly fall below the measured differential cross sections. However, they lie close to the measurements at large values of $Q^2$ and $x$, for backward $\eta^{\gamma}$ and for high values of $E_T^{\gamma}$, where  the LL cross section is expected to be a
substantial fraction of the total.
Also included in Fig.~\ref{fig:xsec3} is the sum of MRST and QQ of
GGP; it gives an improved description of the data over much of the range of the kinematic variables.

\begin{figure}[p]
\vfill
\begin{center}
\epsfig{file=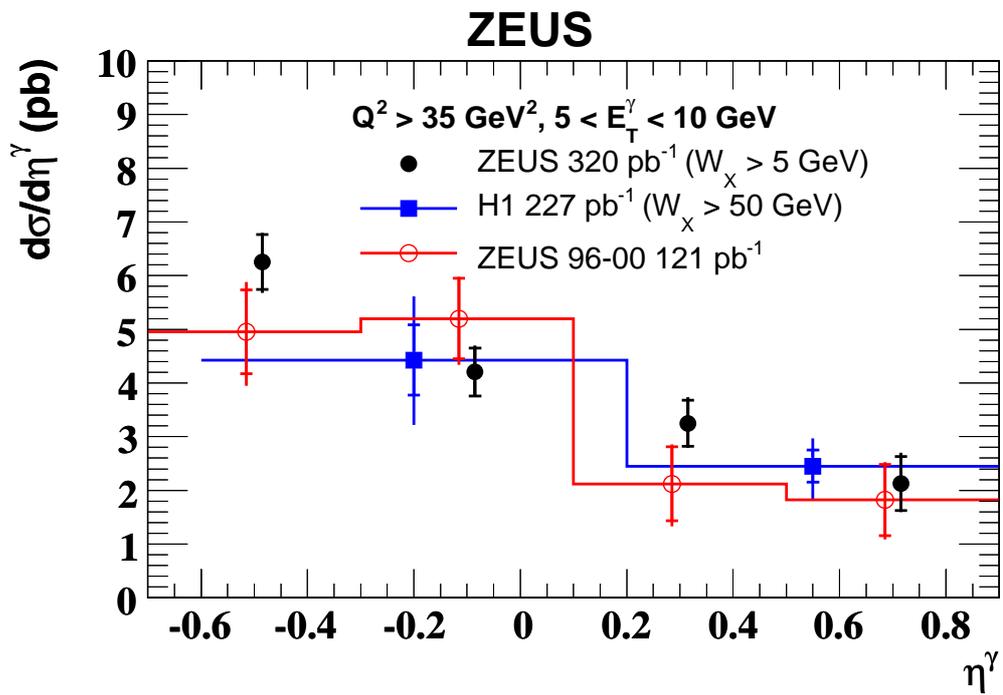,width=14cm}
\end{center}
\caption{\small Isolated photon differential cross-section
  $\frac{d\sigma}{d\eta^{\gamma}}$, compared to previous measurements
at HERA with the additional kinematic restraints $Q^{2} >
35\ \mathrm{GeV}^{2}$ and $5 < E_{T}^{\gamma} < 10\ \mathrm{GeV}$. The histograms show the different binnings used by ZEUS and H1. The symbols are mutually displaced for clarity.
\label{fig:comparison}}
\vfill
\end{figure}

Figure~\ref{fig:comparison} shows the  measured $d\sigma/d\eta^{\gamma}$ compared 
to previous measurements from ZEUS\cite{pl:b595:86} and H1 \cite{epj:c54:371} for 
the restricted range $Q^2>35\,\mathrm{GeV}^2$ and 
$5 < E_T^{\gamma} < 10\,\mathrm{GeV}$. The results are consistent but
the uncertainty in the present measurement is smaller.

\section{Conclusions}

Inclusive isolated photon production has been measured in
deep inelastic scattering using the ZEUS detector at HERA using an integrated luminosity of $320\,\mathrm{pb}^{-1}$.
Differential cross sections as functions of several kinematic variables
are presented for $10< Q^2 < 350\,\mathrm{GeV}^2$ and $W_X>5\,\mathrm{GeV}$
 in the
pseudorapidity range \mbox{$-0.7 < \eta^{\gamma} < 0.9$} for photon 
transverse energies in the range $4 <E_T^{\gamma} < 15
\,\mathrm{GeV}$.  The order $\alpha^3$ predictions of Gehrmann-de Ridder {\it et al.}
 reproduce the shapes of the 
experimental results as functions of transverse energy and pseudorapidity, 
but are lower than the measurements at low
$Q^2$ and low $x$.  The predictions of Martin {\it et al.} mostly fall below the measured
cross sections but are close in the kinematic regions where lepton
emission is expected to be dominant.  An improved description of the data is obtained by appropriately combining the two predictions, suggesting a need for further calculations to exploit the full potential of the measurements.

\section{Acknowledgements}
We appreciate the contributions to the construction and maintenance of the ZEUS detector of many people who are not listed as authors.
The HERA machine group and the DESY computing staff are especially
acknowledged for their success in providing excellent operation of the collider and the data-analysis environment.
We thank the DESY directorate for their strong support and encouragement.  It is a pleasure to thank T.~Gehrmann, W.J.~Stirling and R.~Thorne for useful discussions.

%\Zdetdesc

%\Zctddesc\ZcoosysfnBeta

%\Zmvddesc

%\Zcaldesc

%\vfill\eject

%------------------------------------------------------------------------------
%       Bibliography
%------------------------------------------------------------------------------
%\include{prompt_photons-ref}
\raggedright{
\providecommand{\etal}{et al.\xspace}
\providecommand{\coll}{Collaboration}
\catcode`\@=11
\def\@bibitem#1{%
\ifmc@bstsupport
  \mc@iftail{#1}%
    {;\newline\ignorespaces}%
    {\ifmc@first\else.\fi\orig@bibitem{#1}}
  \mc@firstfalse
\else
  \mc@iftail{#1}%
    {\ignorespaces}%
    {\orig@bibitem{#1}}%
\fi}%
\catcode`\@=12

}
%------------------------------------------------------------------------------
%       Tables
%------------------------------------------------------------------------------
\begin{table}
\begin{center}
\begin{tabular}{|rcr|c|}
\hline
\multicolumn{3}{|c|}{$E_T^{\gamma}$ range}   &        \\[-1.0ex]
\multicolumn{3}{|c|}{ (GeV)}   &  \raisebox{1.5ex}{$\frac{d\sigma}{dE^{\gamma}_T}$ ($\mathrm{pb}\,\mathrm{GeV}^{-1})$}      \\
\hline
4 & -- & 6   & $4.87 \pm  0.28\,\mathrm{(stat.)}\,^{+0.40}_{-0.23}\,\mathrm{(syst.)}$ \\
6 & -- & 8   & $2.40 \pm  0.16\,\mathrm{(stat.)}\,^{+0.09}_{-0.11}\,\mathrm{(syst.)}$ \\
8 & -- & 10  & $1.24 \pm  0.11\,\mathrm{(stat.)}\,^{+0.03}_{-0.04}\,\mathrm{(syst.)}$ \\
10 & -- & 15 & $0.55 \pm  0.04\,\mathrm{(stat.)}\,^{+0.03}_{-0.03}\,\mathrm{(syst.)}$ \\
\hline
\end{tabular}
\end{center}
\caption{Measured differential cross-section $\frac{d\sigma}{dE^{\gamma}_T}$. \label{tab:dsdet}}
\end{table}

\begin{table}
\begin{center}
\begin{tabular}{|rcr|rc|}
\hline
\multicolumn{3}{|c|}{ $\eta^{\gamma}$ range }  & \multicolumn{2}{|c|}{ $\frac{d\sigma}{d\eta^{\gamma}}$ ($\mathrm{pb}$)}      \\
\hline
%-0.7 & -- & -0.3   & $17.44$&$ \pm  1.05\,\mathrm{(stat.)}\,^{+0.45}_{-0.73}\,\mathrm{(syst.)}$ \\
--0.7 & -- & --0.3   & $17.4$&$ \pm  0.9\,\mathrm{(stat.)}\,^{+0.5}_{-0.7}\,\mathrm{(syst.)}$ \\
%-0.3 & -- & 0.1   & $12.97$&$ \pm  0.83\,\mathrm{(stat.)}\,^{+0.64}_{-0.33}\,\mathrm{(syst.)}$ \\
--0.3 & -- & 0.1   & $13.0$&$ \pm  0.8\,\mathrm{(stat.)}\,^{+0.6}_{-0.3}\,\mathrm{(syst.)}$ \\
%0.1 & -- & 0.5  & $10.74$&$ \pm  0.88\,\mathrm{(stat.)}\,^{+0.72}_{-0.43}\,\mathrm{(syst.)}$ \\
0.1 & -- & 0.5  & $10.7$&$ \pm  0.9\,\mathrm{(stat.)}\,^{+0.7}_{-0.4}\,\mathrm{(syst.)}$ \\
%0.5 & -- & 0.9 & $\ 8.65$&$ \pm  0.89\,\mathrm{(stat.)}\,^{+1.08}_{-0.69}\,\mathrm{(syst.)}$ \\
0.5 & -- & 0.9 & $\ 8.7$&$ \pm  0.9\,\mathrm{(stat.)}\,^{+1.1}_{-0.7}\,\mathrm{(syst.)}$ \\
\hline
\end{tabular}
\end{center}
\caption{Measured differential cross-section $\frac{d\sigma}{d\eta^{\gamma}}$.\label{tab:dsdeta}}
\end{table}

\begin{table}
\begin{center}
\begin{tabular}{|rcr|rcrlr|}
\hline
\multicolumn{3}{|c|}{$Q^2$ range}   &  & & & &     \\ [-1.0ex]
\multicolumn{3}{|c|}{ ($\mathrm{GeV}^{2}$)}   & \multicolumn{5}{|c|}{\raisebox{1.5ex}{$\frac{d\sigma}{dQ^2}$ ($\mathrm{pb}\,\mathrm{GeV}^{-2}$)}}     \\
\hline
10 & -- & 20   & $0.414$&$ \pm $&$  0.035\,\mathrm{(stat.)}$&$ ^{+0.045}_{-0.024}$&$\mathrm{(syst.)}$ \\
20 & -- & 40   & $0.279 $&$\pm $&$ 0.020\,\mathrm{(stat.)}$&$^{+0.005}_{-0.014}$&$\mathrm{(syst.)}$ \\
40 & -- & 80  & $0.115$&$ \pm $&$ 0.008\,\mathrm{(stat.)}$&$^{+0.011}_{-0.004}$&$\mathrm{(syst.)}$ \\
80 & -- & 150 & $0.050 $&$ \pm $&$ 0.003\,\mathrm{(stat.)}$&$^{+0.001}_{-0.003}$&$\mathrm{(syst.)}$ \\
150 & -- & 350 & $0.0088 $&$ \pm $&$ 0.0009\,\mathrm{(stat.)}$&$^{+0.0004}_{-0.0003}$&$\mathrm{(syst.)}$\\
\hline
\end{tabular}
\end{center}
\caption{Measured differential cross-section $\frac{d\sigma}{dQ^2}$. \label{tab:dsdq2}}
\end{table}

\begin{table}
\begin{center}
\begin{tabular}{|rcl|rcrlr|}
\hline
\multicolumn{3}{|c|}{$x$ range}    & \multicolumn{5}{|c|}{  $\frac{d\sigma}{dx}$ ($\mathrm{pb}$)}      \\
\hline
0.0002 & -- & 0.001   & $5560$ & $\pm$  & $  380\,\mathrm{(stat.)} $ & $^{+350}_{-250}$  & $\mathrm{(syst.)}$ \\
0.001 & -- & 0.003   & $3920 $ & $ \pm $ & $  230\,\mathrm{(stat.)} $ & $^{+150}_{-180}$  & $\mathrm{(syst.)}$ \\
0.003 & -- & 0.01  & $819 $ & $ \pm $ & $ 58\,\mathrm{(stat.)}$ & $^{+44}_{-42}$  & $\mathrm{(syst.)}$ \\
0.01  & -- & 0.02 & $103 $ & $ \pm $ & $ 16\,\mathrm{(stat.)}$ & $^{+12}_{-16}$  & $\mathrm{(syst.)}$ \\
\hline
\end{tabular}
\end{center}
\caption{Measured differential cross-section $\frac{d\sigma}{dx}$. \label{tab:dsdx}}
\end{table}

%------------------------------------------------------------------------------
%       Figures
%------------------------------------------------------------------------------
%-------------------------------------------------------------------------------
%       Results

%-------------------------------------------------------------------------------

\newpage
\clearpage
%Figure 1

% figure 2

%Figure 3

%Figure 4

%Figure 5

%Figure 7

%
%       ... that's it
%
\end{document}